\documentclass[12pt]{article}
\usepackage{epsf}
\usepackage{graphicx}

\title{ {\bf Semileptonic $B_{s}\rightarrow D_{sJ}(2460)l\nu $ decay in  QCD  }}
\author{\vspace{1cm}\\
 T. M. Aliev \footnote{Permanent address: Institute of Physics, Baku,
 Azerbaijan} \thanks {e-mail:
taliev@metu.edu.tr} ,  K. Azizi \thanks {e-mail:
kazizi@newton.physics.metu.edu.tr} ,  A. Ozpineci \thanks {e-mail:
ozpineci@metu.edu.tr}\\ Physics Department, Middle
East Technical University, Turkey }
 \date{}
\begin{document}
\setlength{\baselineskip}{24pt} \maketitle
\setlength{\baselineskip}{7mm}
\begin{abstract}
 Using three point QCD sum rules method, the form factors relevant to the semileptonic
 $B_{s}\rightarrow D_{sJ}(2460)\ell\nu$ decay are calculated.
The $q^2$ dependences of these form factors
 are evaluated and compared with the heavy quark effective theory predictions. The dependence of the asymmetry parameter
 $\alpha$, characterizing the polarization of $D_{sJ}$ meson, on $q^2$ is studied. The branching ratio
 of this decay is
 also estimated and is shown that it can be easily detected at LHC.
\end{abstract}
\thispagestyle{empty}
\newpage
\setcounter{page}{1}
\section{Introduction}

 Recently, very exciting experimental results have been obtained in charmed hadron spectroscopy. The observation
 of two narrow resonances with charm and strangeness, $D_{sJ}(2317)$ in the $D_{s}\pi^{0}$ invariant mass distribution
\cite{1,2,3,4,5,6}, and $D_{sJ}(2460)$ in the
$D_{s}^{\ast}\pi^{0}$ and $D_{s}\gamma$ mass distribution
\cite{2,3,4,6,7,8}, has raised discussions about the nature of these
states and their quark content \cite{9,10}.
  Analysis of the $D_{s_{0}}(2317)\rightarrow
D_{s}^{\ast}\gamma$, $D_{sJ}(2460)\rightarrow
    D_{s}^{\ast}\gamma$ and $ D_{S_{J}}(2460)\rightarrow D_{s_{0}}(2317)\gamma$ indicates that the quark
    content of these mesons are probably $\overline{c}s$ \cite{11}. In \cite{11} it
    is also shown that finite quark mass effects for the $c$-quark give non-negligible corrections.

    When LHC begins operation, an abundant number of $B_{s}$ mesons will
    be produced creating a real possibility for studying the
    properties of $B_{s}$ meson and its various decay
    channels. One of the possible decay channels  of $B_{s}$ meson
    is its
    semileptonic $B_{s}\rightarrow D_{sJ}(2460)l\nu$ decay.
    Analysis of this decay might yield useful information for understanding the
    structure of the $ D_{sJ}(2460)$ meson.

    It is well known that the
    semileptonic decays of heavy flavored mesons are very promising tools
    for the determination of the elements of the CKM
    matrix, leptonic decay constants as well as the origin of the CP
    violation. In semileptonic decays the long distance dynamics are
    parameterized by transition form factors, calculation of which is
    a central problem for these decays.

    Obviously, for the calculation of the transition form factors,
    nonperturbative approaches are needed. Among the nonperturbative
    approaches, the QCD sum rules method \cite{12} received special
    attention, because this method is based on the fundamental QCD
    Lagrangian. This method has been successfully applied to a wide
    variety of problems in hadron physics(for a review see \cite{13}). The
    semileptonic decay $D\rightarrow \overline{K}^{0}l\nu$ is
    studied using the QCD sum rules with three point correlation
    function in \cite{14}. Then, the semileptonic decays $D^{+}\rightarrow
    K^{0\ast}e^{+}\nu~e$ \cite{15}, $D\rightarrow \pi~l\nu$ \cite{16}, $D\rightarrow
    \rho~l\overline{\nu}e$ \cite{17} and $B\rightarrow
    D(D^{\ast})l\nu~e$ \cite{18} are studied in the same framework.
    In the
    present work we study the semileptonic decay of $B_{s}$ meson
    to positive parity $D_{sJ}(2460)$ meson,i.e, $B_{s}\rightarrow
    D_{sJ}(2460)\ell\nu$, within QCD sum rules method. Note that, in \cite{19},
    the decay $B_{s}\rightarrow D_{s_{0}}(2317)l\nu$ has been studied using the QCD sum rules.

    The paper is organized
    as follows: In section  II  the sum rules for the transition form factors are
    calculated; section  III  is devoted to the numerical analysis, discussion and our conclusions.

\section{Sum rules for the $B_{s}\rightarrow D_{sJ}(2460)\ell\nu$ transition form factors }
The $B_s \rightarrow  D_{sJ}$ transition proceeds by the $b\rightarrow c$ transition at the quark level.
The matrix element for the quark level process can be written as:
\begin{equation}\label{lelement}
M_{q}=\frac{G_{F}}{\sqrt{2}} V_{cb}~\overline{\nu}
~\gamma_{\mu}(1-\gamma_{5})l~\overline{c}
~\gamma_{\mu}(1-\gamma_{5}) b \\
\end{equation}
In order to obtain the matrix elements for $B_{s}\rightarrow
D_{sJ}(2460)\ell\nu$ decay, we need to sandwich Eq. (\ref{lelement})
between initial and final meson states. So, the amplitude of the
$B_{s}\rightarrow D_{sJ}(2460)\ell\nu$ decay can be written as:
\begin{equation}\label{2au}
M=\frac{G_{F}}{\sqrt{2}} V_{cb}~\overline{\nu}
~\gamma_{\mu}(1-\gamma_{5})l<D_{sJ}\mid~\overline{c}
~\gamma_{\mu}(1-\gamma_{5}) b\mid B_{s}>
\end{equation}
 The main problem is the calculation of the matrix element
$<D_{sJ}\mid\overline{c}\gamma_{\mu}(1-\gamma_{5}) b\mid B_{s}>$ appearing in Eq. (\ref{2au}).
Both vector and axial vector part of
  $~\overline{c}~\gamma_{\mu}(1-\gamma_{5}) b~$  contribute to the
matrix element considered above. From Lorentz invariance and
parity considerations, this matrix element can
be parameterized in terms of the form factors in the following way:
\begin{equation}\label{3au}
<D_{sJ}(p',\varepsilon)\mid\overline{c}\gamma_{\mu}\gamma_{5}
b\mid
B_s(p)>=\frac{f_{V}(q^2)}{(m_{B_{s}}+m_{D_{sJ}})}\varepsilon_{\mu\nu\alpha\beta}
\varepsilon^{\ast\nu}p^\alpha p'^\beta
\end{equation}
\begin{eqnarray}\label{4au}
< D_{sJ}(p',\varepsilon)\mid\overline{c}\gamma_{\mu} b\mid
B_{s}(p)> &=&i\left[f_{0}(q^2)(m_{B_{s}}
+m_{D_{sJ}})\varepsilon_{\mu}^{\ast}
\right. \nonumber \\
+ 
\frac{f_{+}(q^2)}{(m_{B_{s}}+m_{D_{sJ}})}(\varepsilon^{\ast}p)P_{\mu}
&+& \left. \frac{f_-(q^2)}{(m_{B_{s}}+m_{D_{sJ}})}(\varepsilon^{\ast}p)q_{\mu}\right]
\end{eqnarray}
where $f_{V}(q^2)$, $f_{0}(q^2)$, $f_{+}(q^2)$ and $f_{-}(q^2)$ are
the transition form factors and $P_{\mu}=(p+p')_{\mu}$,
$q_{\mu}=(p-p')_{\mu}$. In all following discussions, for
customary, we will use following redefinitions:
\begin{eqnarray}\label{eq5}
f_{V}'(q^2)&=&\frac{f_{V}(q^2)}{(m_{B_{s}}+m_{D_{sJ}})}~,~~~~~~~~~~~~f_{0}'(q^2)=f_{0}(q^2)(m_{B_{s}}
+m_{D_{sJ}})\nonumber
\\
f_{+}'(q^2)&=&\frac{f_{+}(q^2)}{(m_{B_{s}}+m_{D_{sJ}})}~,~~~~~~~~~~~~
f_{-}'(q^2)=\frac{f_{- }(q^2)}{(m_{B_{s}}+m_{D_{sJ}})}
\end{eqnarray}
For the calculation of these  form factors,  QCD sum rules
method will be employed. We start by considering the following correlator:
\begin{equation}\label{6au}
\Pi _{\mu\nu}^{V;A}(p^2,p'^2,q^2)=i^2\int
d^{4}xd^4ye^{-ipx}e^{ip'y}<0\mid T[J _{\nu D_{sJ}}(y)
J_{\mu}^{V;A}(0) J_{B_{s}}(x)]\mid  0>
\end{equation}
where $J _{\nu D_{sJ}}(y)=\overline{s}\gamma_{\nu} \gamma_{5}c$,
$J_{B_{s}}(x)=\overline{b}\gamma_{5}s$ ,
 $J_{\mu}^{V}=~\overline{c}\gamma_{\mu}b $ and $J_{\mu}^{A}=~\overline{c}\gamma_{\mu}\gamma_{5}b$
 are the interpolating currents of the  $D_{sJ}$, $B_{s} $,
 vector and axial vector
currents respectively.

To calculate the phenomenological part of the correlator
given in Eq. (\ref{6au}), two
complete sets of intermediate states with the same quantum number as
the currents $J_{D_{sJ}}$ and $J_{B_{s}}$ respectively are inserted. As a result of this
procedure we get the following
representation of the above-mentioned correlator:
\begin{eqnarray} \label{7au}
&&\Pi _{\mu\nu}^{V,A}(p^2,p'^2,q^2)=
\nonumber \\
&& \frac{<0\mid J_{D_{sJ}}^{\nu}
\mid D_{sJ}(p')^{\varepsilon}><D_{sJ}(p')^{\varepsilon}\mid
J_{\mu}^{V,A}\mid B_{s}(p)><B_{s}(p)\mid J_{Bs}\mid
0>}{(p'^2-m_{D_{sJ}}^2)(p^2-m_{Bs}^2)}+\cdots
\nonumber \\
\end{eqnarray} 
 where $\cdots$ represent contributions coming from higher states and continuum. The matrix
 elements in Eq. (\ref{7au}) are defined in the standard way as:
\begin{equation}\label{8au}
 <0\mid J^{\nu}_{D_{sJ}} \mid
D_{sJ}(p')>=f_{D_{sJ}}m_{D_{sJ}}\varepsilon^{\nu}~,~~<B_{s}(p)\mid
J_{Bs}\mid 0>=-i\frac{f_{B_{s}}m_{B_{s}}^2}{m_{b}+m_{s}}
\end{equation}
where $f_{D_{sJ}}$ and $f_{B_{s}}$  are the leptonic decay
constants of $D_{sJ} $ and $B_{s}$ mesons, respectively. Using
Eq. (\ref{3au}), Eq. (\ref{4au}) and Eq. (\ref{8au})
and performing summation over the polarization of the $D_{sJ}$ meson, Eq. (\ref{7au}) can be written as:
\begin{eqnarray}\label{9amplitude}
\Pi_{\mu\nu}^{V}(p^2,p'^2,q^2)&=&-\frac{f_{B_{s}}m_{B_{s}}^2}{(m_{b}+m_{s})}\frac{f_{D_{sJ}}m_{D_{sJ}}}
{(p'^2-m_{D_{sJ}}^2)(p^2-m_{Bs}^2)} \times
[f_{0}'g_{\mu\nu}+f_{+}'P_{\mu}p_{\nu} \nonumber
\\ &+&f_{-}'q_{\mu}p_{\nu}]+
\mbox{excited states.}\nonumber
\\\Pi_{\mu\nu}^{A}(p^2,p'^2,q^2)&=&
-i\varepsilon_{\mu\nu\alpha\beta}p'^{\alpha}p^{\beta}\frac{f_{B_{s}}m_{B_{s}}^2}{(m_{b}+m_{s})}\frac{f_{D_{sJ}}m_{D_{sJ}}}
{(p'^2-m_{D_{sJ}}^2)(p^2-m_{Bs}^2)}f_{V}' +
\nonumber \\
&&\mbox{excited states.}
\end{eqnarray}

In accordance with the QCD sum rules philosophy, $\Pi
_{\mu\nu}(p^2,p'^2,q^2)$ can also be calculated from QCD side with the
help of the operator product expansion(OPE) in the deep
Euclidean region  $p^2 \ll (m_{b}+m_{c})^2 $ and $p'^2 \ll
(m_{c}+m_{s})^2$. 
The theoretical part of the correlator is calculated
by means of OPE, and up to operators having dimension $d=6$, it is
determined by the bare-loop and the power corrections from the
operators with $d=3$, $<\overline{\psi}\psi>$,
$d=4$, $m_{s}<\overline{\psi}\psi>$,
$d=5$, $m_{0}^{2}<\overline{\psi}\psi>$ and
$d=6$, $<\overline{\psi}\psi\bar \psi \psi>$. In calculating the
$d=6$ operator, vacuum saturation approximation is used to set $<\overline{\psi}\psi\bar \psi \psi> = <\overline{\psi}\psi>^2$.
 In calculating the
bare-loop contribution, we first write the double dispersion
representation for the coefficients of corresponding Lorentz
structures appearing in the correlation function as:
\begin{equation}\label{10au}
\Pi_i^{'per}=-\frac{1}{(2\pi)^2}\int
dsds'\frac{\rho_{i}(s,s',q^2)}{(s-p^2)(s'-p'^2)}+\textrm{
subtraction terms}
\end{equation}
The spectral densities $\rho_{i}(s,s',q^2)$ can be calculated from
the usual Feynman integral with the help of Cutkosky rules, i.e. by
replacing the quark propagators with Dirac delta functions:
$\frac{1}{p^2-m^2}\rightarrow-2\pi\delta(p^2-m^2),$ which implies
that all quarks are real. After standard calculations for the
corresponding spectral densities we obtain:
\begin{eqnarray}\label{11au}
\rho_{V}(s,s',q^2)&=&N_{c}I_{0}(s,s',q^2)\left[{m_{s}+(m_{s}-m_{b})B_{1}+(m_{s}+m_{c})B_{2}}\right],\nonumber\\
\rho_{0}(s,s',q^2)&=&N_{c}I_{0}(s,s',q^2)[8(m_{b}-m_{s})A_{1}-4m_{b}m_{c}m_{s}\nonumber\\&+&
4(m_{s}- m_{b}+m_{c
})m_{s}^2-2(m_{s}+m_{c})(\Delta+m_{s}^2)\nonumber
\\&-&
2(m_{s}-m_{b})(\Delta'+m_{s}^2)+2m_{s}u]\nonumber \\
\rho_{+}(s,s',q^2)&=&N_{c}I_{0}(s,s',q^2)[4(m_{b}-m_{s})(A_{2}+A_{3})+2(m_{b}-3m_{s})B_{1}
\nonumber \\
&& -2(m_{c}+m_{s})B_{2}-2m_{s}]
,\nonumber \\
\rho_{-}(s,s',q^2)&=&N_{c}I_{0}(s,s',q^2)[4(m_{b}-m_{s})(A_{2}-A_{3})-2(m_{b}+m_{s})B_{1}
\nonumber \\
&& +2(m_{c}+m_{s})B_{2} +2m_{s}]\nonumber \\
\end{eqnarray}
where
\begin{eqnarray}\label{12}
I_{0}(s,s',q^2)&=&\frac{1}{4\lambda^{1/2}(s,s',q^2)},\nonumber\\
 \lambda(s,s',q^2)&=&s^2+s'^2+q^4-2sq^2-2s'q^2-2ss',\nonumber \\
\Delta'&=&(s'-m_{c}^2 + m_{s}^2),\nonumber\\
\Delta&= &(s-m_{b}^2 +
m_{s}^2),\nonumber\\
 u &=& s + s' - q^2,\nonumber\\
 B_{1}&=&\frac{1}{\lambda(s,s',q^2)}[2s'\Delta-\Delta'u],\nonumber\\
 B_{2}&=&\frac{1}{\lambda(s,s',q^2)}[2s\Delta'-\Delta u],\nonumber\\
 A_{1}&=&\frac{1}{2\lambda(s,s',q^2)}[\Delta'^{2}s+2\Delta'm_{s}^2s+m_{s}^4s+\Delta^2s'+2 \Delta m_{s}^2s'
 \nonumber \\
 && +m_{s}^4s'- 4m_{s}^2ss'-\Delta\Delta'u-\Delta m_{s}^2u-\Delta'm_{s}^2u-m_{s}^4u+m_{s}^2u^2],\nonumber\\
 A_{2}&=&\frac{1}{\lambda^{2}(s,s',q^2)}[2\Delta'^2ss'+4\Delta'm_{s}^2ss'+2m_{s}^4ss'+6\Delta^2s'^2
 \nonumber \\
 && +12\Delta m_{s}^2s'^2 +6m_{s}^4s'^2-8m_{s}^2ss'^2-6\Delta\Delta's'u
 \nonumber \\
 && -6\Delta m_{s}^2s'u-6\Delta'm_{s}^2s'u-6m_{s}^4s'u+\Delta'^2u^2+2\Delta'
 m_{s}^2u^2
 \nonumber\\
 &&+m_{s}^4u^2+2m_{s}^2s'u^2],\nonumber\\
 A_{3}&=&\frac{1}{\lambda^{2}(s,s',q^2)}[4\Delta\Delta'ss'+4\Delta m_{s}^2ss'+4\Delta'm_{s}^2ss'+4m_{s}^4ss'
 \nonumber \\
 &&-3\Delta'^2su- 6\Delta'm_{s}^2su -3m_{s}^4su-3\Delta^2s'u-6\Delta m_{s}^2s'u
 \nonumber \\
 &&-3m_{s}^4s'u+4m_{s}^2ss'u+2\Delta\Delta'u^2+
 2\Delta m_{s}^2u^2+2\Delta' m_{s}^2u^2
 \nonumber \\
 &&+2m_{s}^4u^2-m_{s}^2u^3]\nonumber\\
 \end{eqnarray}
 The subscripts V, 0 and $\pm$ correspond to the coefficients of the
 structures proportional to $i\varepsilon_{\mu\nu\alpha\beta}p'^{\alpha}p^{\beta}$, $g_{\mu\nu}$ and $\frac{1}{2}(p_{\mu}p_{\nu}
 \pm p'_{\mu}p_{\nu})$ respectively. In Eq. (\ref{11au}) $N_{c}=3$ is the number of colors.

 The integration region for the perturbative contribution
 in Eq. (\ref{10au}) is determined from the condition that arguments of the
 three $\delta$ functions must vanish simultaneously. The physical
 region in s and s' plane is described by the following
 inequalities:\\
 \begin{equation}\label{13au}
 -1\leq\frac{2ss'+(s+s'-q^2)(m_{b}^2-s-m_{s}^2)+(m_{s}^2-m_{c}^2)2s}{\lambda^{1/2}(m_{b}^2,s,m_{s}^2)\lambda^{1/2}(s,s',q^2)}\leq+1
\end{equation}

For the contribution of power corrections, i.e. the contributions of operators with
dimensions $d=3$, $4$ and $5$, we obtain the following results:
\begin{eqnarray}\label{14au}
f_{V}^{'(3)}+f_{V}^{'(4)}+f_{V}^{'(5)}&=&\frac{1}{rr'}<\overline{s}s>-\frac{m_{s}}{2}<\overline{s}s>[\frac{-m_{c}}{rr'^2}+\frac{m_{b}}{r'r^2}]
\nonumber \\ && +\frac{m_{s}^2}{2}
<\overline{s}s>
[\frac{2m_{c}^2}{r'^3r}+\frac{m_{b}^2+m_{c}^2-q^2}{r'^2r^2}+\frac{2m_{b}^2}{r'r^3}]
\nonumber \\
&& -\frac{m_{0}^2}{6}<\overline{s}s>
[\frac{3m_{c}^2}{r'^3r}+\frac{3m_{b}^2}{r'r^3}+\frac{2}{r'r^2}i
\nonumber \\ &&
+\frac{2m_{b}^2+2m_{c}^2+m_{b}m_{c}
-2q^2}{r'^2r^2}]
\nonumber \\
f_{0}^{'(3)}+f_{0}^{'(4)}+f_{0}^{'(5)}&=&\frac{(m_{b}-m_{c})^2-q^2}{2rr'}<\overline{s}s>
\nonumber \\ &&
+\frac{m_{s}}{4}
<\overline{s}s>[\frac{-2m_{b}m_{c}^2+m_{c}m_{b}^{2}+m_{c}^3-m_{c}q^2}{rr'^2}\nonumber\\&&-\frac{m_{c}+m_{b}}{rr'}
+\frac{2m_{c}m_{b}^2-m_{b}^3-m_{b}m_{c}^2+m_{b}q ^2}{r'r^2}]\nonumber \\
&&+\frac{m_{s}^2}{16}<\overline{s}s>\left\{\frac{-16m_{b}m_{c}^3+8m_{c}^2m_{b}^2+8m_{c}^4-8m_{c}^2q^2}{r'^3r} \right.
\nonumber\\
&&+\frac{-16m_{b}^3m_{c}
+8m_{b}^4+8m_{c}^2m_{b}^2-8m_{b}^2q^2}{r'r^3}\nonumber \\
&&+\frac{4m_{c}^2-8m_{b}m_{c}+4m_{b}^{2}-4q^2}{r'^2r}
\nonumber \\
&& +\frac{4m_{c}^2-8m_{b}m_{c}+4m_{b}^{2}-4q^2}{r'r^2}-\frac{8}{r'r}
 \nonumber
\\&&+\frac{1}{r'^2 r^2} \left[-8m_{b}^3m_{c}-8m_{b}m_{c}^3+8m_{b}m_{c}q^2+4m_{b}^{4} \right.
\nonumber \\ &&
+\left. \left. 8m_{c}^2m_{b}^2 +4m_{c}^4-8m_{b}^2q^2-8m_{c}^2q^2+4q^4\right]\right\} 
\nonumber \\
&&-\frac{m_{0}^2}{12}<\overline{s}s>[\frac{3m_{c}^2(m_{c}^2+m_{b}^2-2m_{b}m_{c}-q^2)}{r'^3r}
\nonumber \\
&&+ 3 m_b^2\frac{m_{c}^2
+m_{b}^2-2m_{b}m_{c}-q^2}{r'r^3}\nonumber \\&&+
\frac{-3m_{b}m_{c}(m_{c}^2+m_{b}^2-q^2)+2(m_{c}^2+m_{b}^2-q^2)^2-2m_{c}^2m_{b}^2}{r'^2r^2}\nonumber\\&&+\frac{3m_{c}
(m_{c}-m_{b}) +2(m_{b}^2-q^2)}{rr'^2}
\nonumber \\
&& +\frac{3m_{b}(-3m_{c}+m_{b})+4(m_{c}^2-q^2)}{r^2r'}-\frac{2}{rr'}]\nonumber \\
f_{+}^{'(3)}+f_{+}^{'(4)}+f_{+}^{'(5)}&=&-\frac{1}{2rr'}<\overline{s}s>+\frac{m_{s}}{4}<\overline{s}s>[\frac{-m_{c}}{rr'^2}+\frac{m_{b}}{r'r^2}]
\nonumber \\
&&+\frac{m_{s}^2}{32}
<\overline{s}s>
[-\frac{16m_{c}^2}{r'^3r}-\frac{16m_{b}^2}{r'r^3}
+\frac{16}{r'r^2}+\frac{-8m_{b}^2-8m_{c}^2+8q^2}{r^2r'^2}]
\nonumber \\ &&
+\frac{m_{0}^2}{12}<\overline{s}s>[\frac{3m_{c}^2}{r'^3r}+\frac{3m_{b}^2}{r'r^3}-
\frac{2}{r'r^2}\nonumber \\&&+\frac{2m_{b}^2+2m_{c}^2+m_{b}m_{c}-2q^2}{r'^2r^2}]\nonumber \\
f_{-}^{'(3)}+f_{-}^{'(4)}+f_{-}^{'(5)}&=&\frac{1}{2rr'}<\overline{s}s>-\frac{m_{s}}{4}<\overline{s}s>[\frac{-m_{c}}{rr'^2}+\frac{m_{b}}{r'r^2}]
\nonumber \\ &&
+\frac{m_{s}^2}{32}
<\overline{s}s>
[\frac{16m_{c}^2}{r'^3r}+\frac{16m_{b}^2}{r'r^3}
+\frac{16}{r'r^2}+\frac{8m_{b}^2+8m_{c}^2-8q^2}{r^2r'^2}]
\nonumber \\ && 
-\frac{m_{0}^2}{12}<\overline{s}s>[\frac{3m_{c}^2}{r'^3r}+\frac{3m_{b}^2}{r'r^3}+
\frac{6}{r'r^2}\nonumber
\\&&+\frac{2m_{b}^2+2m_{c}^2+m_{b}m_{c}-2q^2}{r'^2r^2}]
\end{eqnarray}
where $r=p^2-m_{b}^2,r'=p'^2-m_{c}^2$. We would like to note that
the contributions of operators with $d=6$ are also calculated.
Numerically their contributions
to the corresponding sum rules turned out to be very small and therefore we did
not present their explicit expressions. Note also that, in the
present work we neglect the $\alpha_{s}$ corrections to the bare
loop. For consistency, we also neglect $\alpha_{s}$ corrections in
determination of the leptonic decay constants $f_{B_{s}}$ and
$f_{D_{sJ}}$. 

The QCD sum rules for the form
factors $f'_{V}$, $f'_{0}$, $f'_{+}$ and $f'_{-}$ 
is obtained by equating the phenomenological expression
given in Eq. (\ref{9amplitude}) and the OPE expression given by Eqs.
(\ref{11au}-\ref{14au}) and 
applying double Borel transformations with
respect to the variables $p^2$ and $p'^2$ ($p^2\rightarrow
M_{1}^2,p'^2\rightarrow M_{2}^2$) in order to suppress the
contributions of higher states and continuum: 
\begin{eqnarray}\label{15au}
f'_{i}(q^2)=-\frac{(m_{b}+m_{s})
}{f_{B_{s}}m_{B_{s}}^2}\frac{1}{f_{D_{sJ}}m_{D_{sJ}}}e^{m_{B_{s}}^2/M_{1}^2+m_{D_{sJ}}^2/M_{2}^2}
\nonumber
\\\times[-\frac{1}{(2\pi)^2)}\int_{(m_b+m_s)^2}^{s_0} ds \int_{(m_c+m_s)^2}^{s_0'} ds'\rho_{i}(s,s',q^2)e^{-s/M_{1}^2-s'/M_{2}^2}\nonumber
\\+\hat{B}(f_{i}^{(3)}+f_{i}^{(4)}+f_{i}^{(5)})]\nonumber\\
\end{eqnarray}
where $i=V,0$ and $\pm$, and $\hat B$ denotes the double Borel
transformation operator. In Eq. (\ref{15au}), in order to subtract the contributions of the
higher states and the continuum, quark-hadron duality assumption is used, i.e. it is assumed that
\begin{eqnarray}
\rho^{higher states}(s,s') = \rho^{OPE}(s,s') \theta(s-s_0) \theta(s-s'_0)
\end{eqnarray}
In calculations the following rule
for double Borel
transformations is used:\\
\begin{equation}\label{16au}
\hat{B}\frac{1}{r^m}\frac{1}{r'^n}\rightarrow(-1)^{m+n}\frac{1}{\Gamma(m)}\frac{1}{\Gamma
(n)}e^{-s/m_{b}^2}e^{-s'/m_{c}^2}\frac{1}{(M^{2})^{m-1}(M'^{2})^{n-1}}.
\end{equation}
%
\section{Numerical analysis}
In this section we present our numerical analysis for the
form factors $f_{V}(q^2)$, $f_{0}(q^2)$, $f_{+}(q^2)$ and
$f_{-}(q^2)$. From sum rule expressions of these form factors we see
that  the condensates, leptonic decay constants of $B_{s}$ and
$D_{sJ}$ mesons, continuum thresholds $s_{0}$ and  $s'_{0} $ and
Borel parameters $M_{1}^2$ and $M_{2}^2$ are the main input
parameters. In further numerical analysis we choose the value of the
condensates at a fixed renormalization scale of about $1$ GeV.
The values of the condensates are\cite{20}
:
$<\overline{\psi}\psi\mid_{\mu=1~GeV}>=-(240\pm10~MeV)^3$,
$<\overline{s}s>=(0.8\pm0.2)<\overline{\psi}\psi>$ and $m_{0}^2=0.8~GeV^2$.
 The quark masses are taken to be $ m_{c}(\mu=m_{c})=
 1.275\pm
 0.015~ GeV$, $m_{s}(1~ GeV)\simeq 142 ~MeV$ \cite{21} and $m_{b} =
(4.7\pm
 0.1)~GeV$ \cite{20} also the mesons masses are taken to be $m_{D_{sJ}}=2.46~GeV$ and $ m_{B_{s}}=5.3~GeV$. For
 the values of the leptonic decay
constants of $B_{s}$ and $D_{sJ} $ mesons we use the results
obtained from two-point QCD analysis: $f_{B_{s}} = 209\pm
 38~ MeV $ \cite{13} and $f_{D_{sJ}} =225\pm25
  ~MeV $\cite{11}. The threshold parameters
$s_{0}$ and $s_{0}' $ are also determined from the two-point QCD sum
rules: $s_{0} =(35\pm 2)~ GeV^2$ \cite{12} and $s_{0}' =9~ GeV^2 $ \cite{11}.
The Borel parameters $M_{1}^2$ and $M_{2}^2 $ are auxiliary
quantities and therefore the results of physical quantities should
not depend on them. In
QCD sum rule method, OPE is truncated at some finite order, leaving a residual dependence
on the Borel parameters. For this
reason, working regions for the Borel
parameters should be chosen such that in these regions  form factors are practically
independent of them. The working regions for the Borel
parameters $M_{1}^2 $ and $M_{2}^2$ can be determined by requiring that, on the one
side, the continuum contribution should be
small, and on the other side, the contribution of the operator with the highest dimension
should be small. As a result of the above-mentioned requirements, the working regions
are determined to be $ 10~ GeV^2 < M_{1}^2 <20~ GeV^2 $ and $
4~ GeV^2 <M_{2}^2 <10 ~GeV^2$.

\begin{figure}
\vspace*{-1cm}
\begin{center}
\includegraphics[width=10cm]{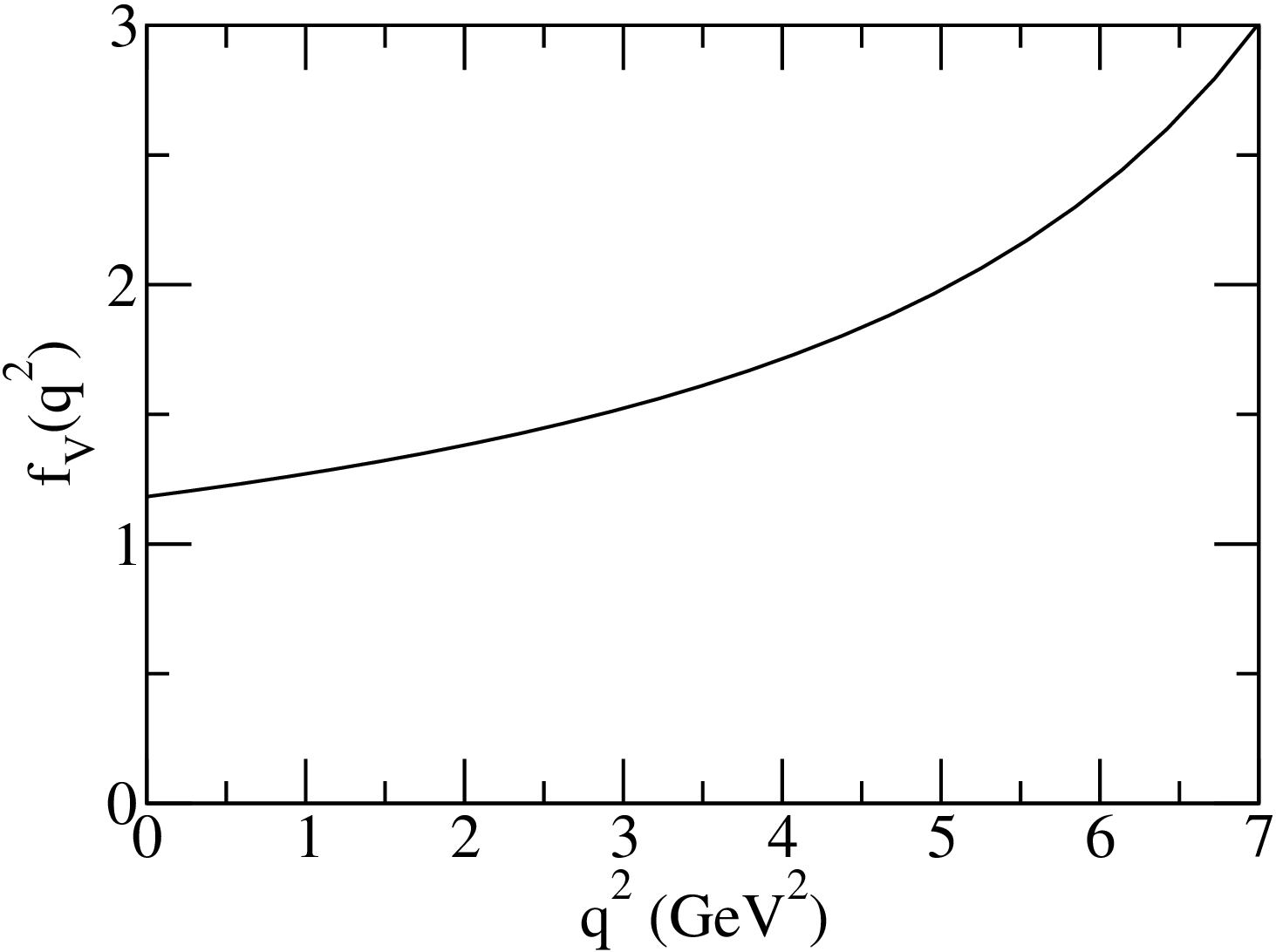}
\end{center}
\caption{The dependence of $f_{V}$ on
 $q^2$ at $M_{1}^2=15~GeV^2$, $M_{2}^2=6~GeV^2$, $s_{0}=35~GeV^2$ and $s_{0}'=9~GeV^2$. } \label{fig1}
\end{figure}

\begin{figure}
\begin{center}
\includegraphics[width=10cm]{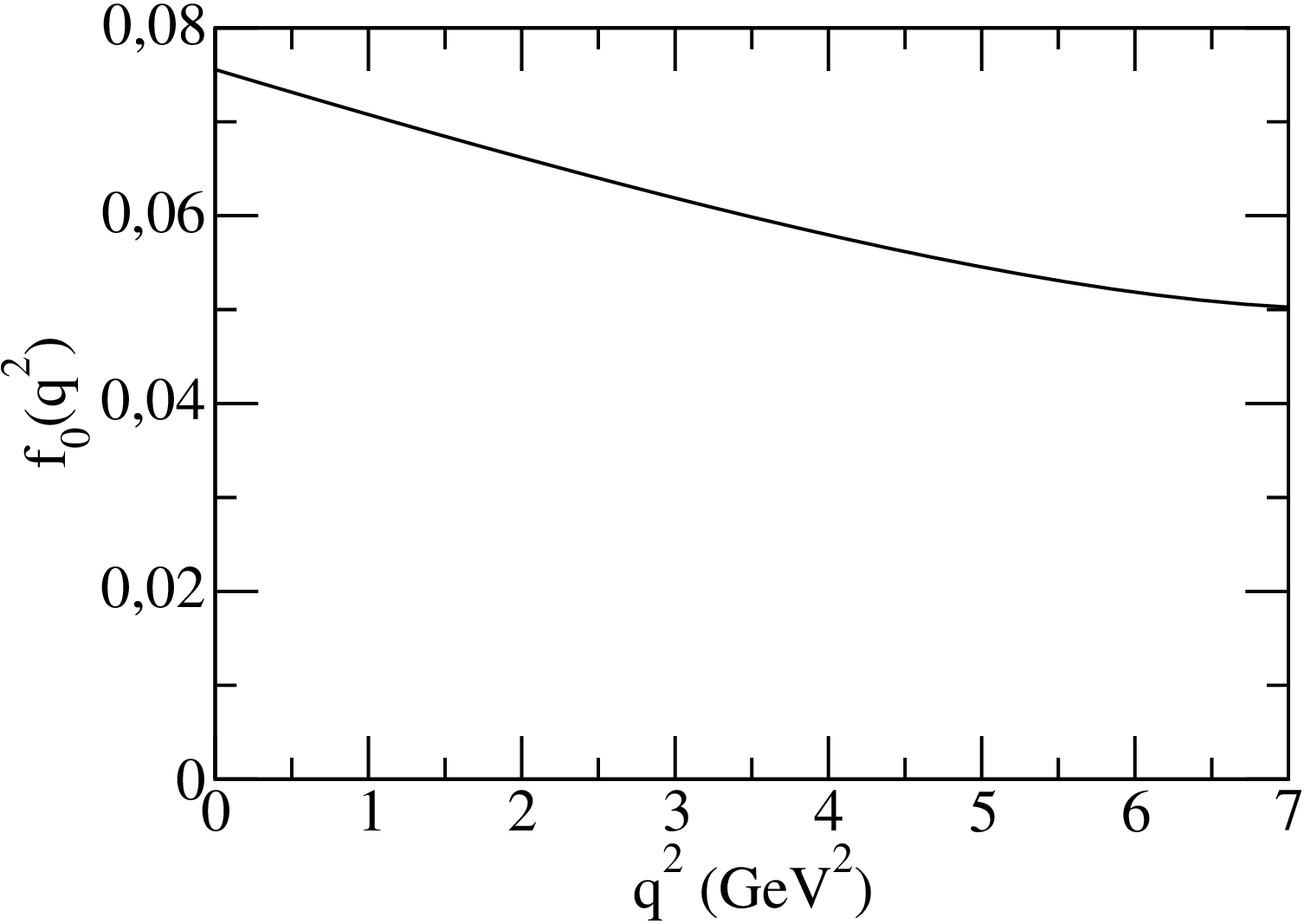}
\end{center}
\caption{ The dependence of $f_{0}$ on
 $q^2$ at $M_{1}^2=15~GeV^2$, $M_{2}^2=6~GeV^2$, $s_{0}=35~GeV^2$ and $s_{0}'=9~GeV^2$.} \label{fig2}
\end{figure}

\begin{figure}
\vspace{-2cm}
\begin{center}
\includegraphics[width=10cm]{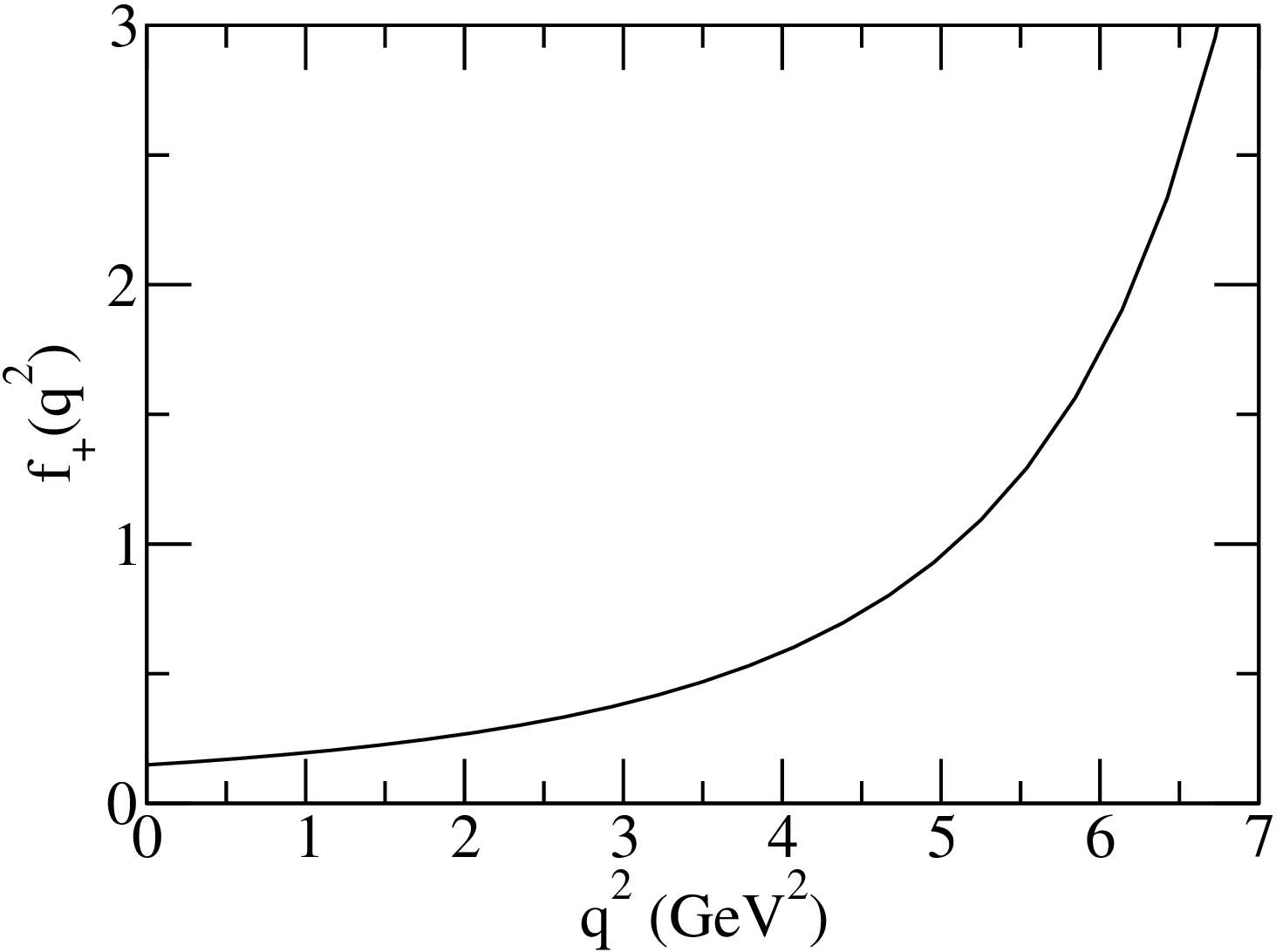}
\end{center}
\caption{The dependence of $f_{+}$ on
 $q^2$ at $M_{1}^2=15~GeV^2$, $M_{2}^2=6~GeV^2$, $s_{0}=35~GeV^2$ and $s_{0}'=9~GeV^2$.} \label{fig3}
\end{figure}

\begin{figure}
\begin{center}
\includegraphics[width=10cm]{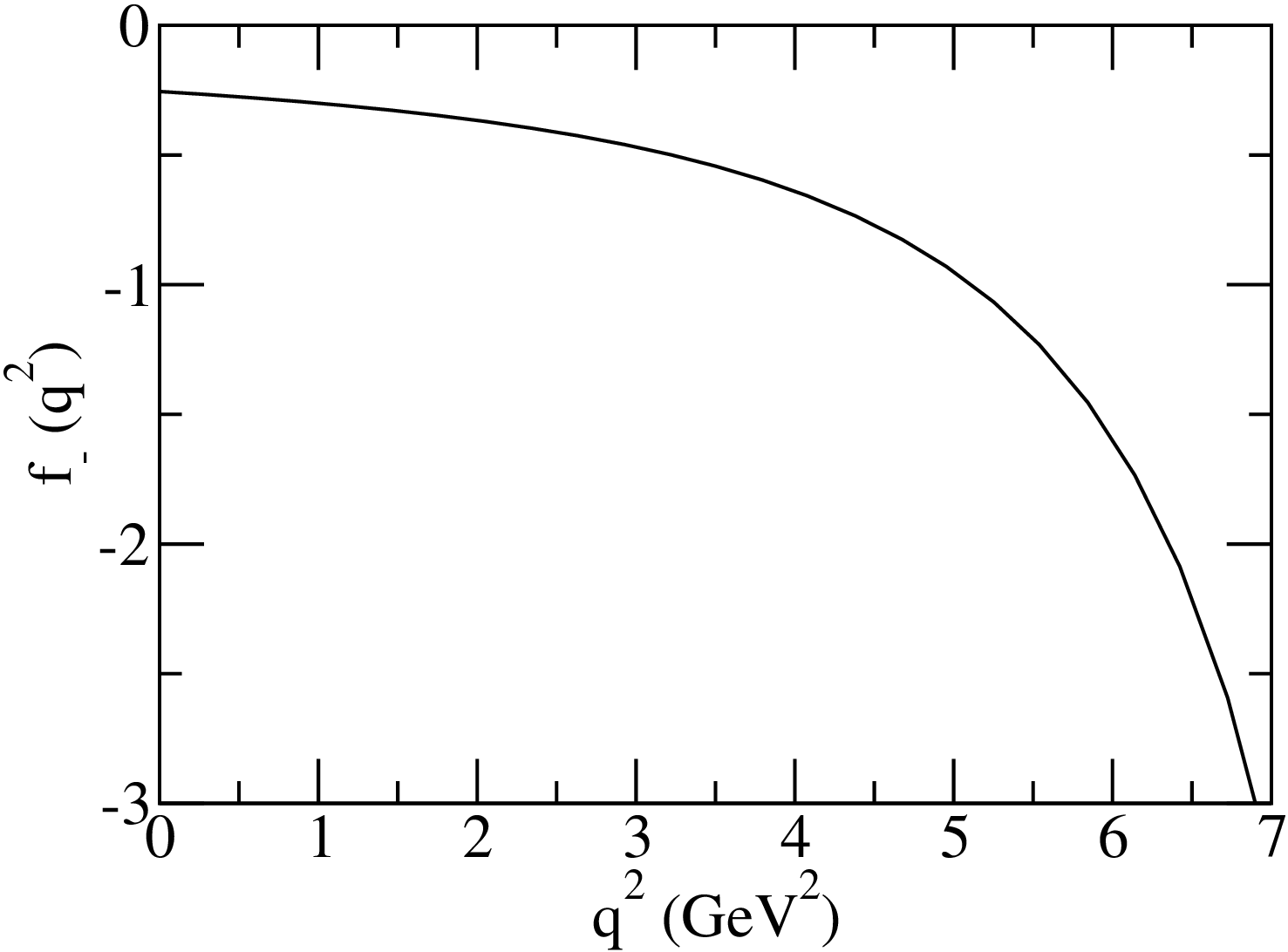}
\end{center}
\caption{The dependence of $f_{-}$ on
 $q^2$ at $M_{1}^2=15~GeV^2$, $M_{2}^2=6~GeV^2$, $s_{0}=35~GeV^2$ and $s_{0}'=9~GeV^2$.} \label{fig4}
\end{figure}

 In order to estimate the width of $B_{s} \rightarrow D_{sJ}l\nu$ it is necessary to know
 the $q^2$ dependence of the form factors $ f_{V}(q^2)$, $f_{0}(q^2)$, $f_{+}(q^2)$ and $f_{-}(q^2)$ in the whole
physical region $ m_{l}^2 \leq q^2 \leq (m_{B_{s}} -
m_{D_{sJ}})^2$. The $q^2 $ dependence of the form factors can be
calculated from QCD sum rules (for details, see  \cite{15,16}). For
extracting the $q^2$ dependence of the form factors from QCD sum
rules we should consider a range $ q^2$ where the correlator
function can reliably be calculated. For this purpose we have to
stay approximately $1~ GeV^2$ below the perturbative cut, i.e., up
to $q^2 =8 ~GeV^2$. In order to extend our results to the full
physical region, we look for parameterization of the form factors in
such a way that in the region $0 \leq q^2 \leq 8~ GeV^2$, this
parameterization coincides with the sum rules prediction. The
dependence of form factors $f_{V}(q^2)$, $f_{0}(q^2)$, $f_{+}(q^2)$
and $f_{-}(q^2)$  on $q^2$ are given in Figs.\ref{fig1}, \ref{fig2},
\ref{fig3} and \ref{fig4}, respectively. Our numerical calculations
shows that the best parameterization of the
form factors with respect to $q^2$ are as follows:\\
 \begin{equation}\label{17au}
 f_{i}(q^2)=\frac{f_{i}(0)}{1+\tilde \alpha\hat{q}+\tilde \beta\hat{q}^2+\tilde \gamma\hat{q}^3+\tilde \lambda\hat{q}^4}
\end{equation}
where $\hat{q}=q^2/m_{B_{s}}^2$. The values of the parameters
 $f_{i}(0),\tilde \alpha,\tilde \beta,\tilde \gamma$, and $\tilde \lambda$ are
given in the Table 1.

\begin{table}[h]
\centering
\begin{tabular}{|c|c|c|c|c|c|} \hline
  & f(0)  & $\tilde \alpha$ & $\tilde \beta$& $\tilde \gamma$& $\tilde \lambda$\\\cline{1-6}
 $f_{V}$ & 1.18 & -1.87 & -1.88& -2.41& 3.34\\\cline{1-6}
 $f_{0}$ & 0.076  & 1.85 & 0.89& 19.0& -79.3\\\cline{1-6}
 $f_{+}$ & 0.13  & -7.14 & 11.6& 21.3& -59.8\\\cline{1-6}
 $f_{-}$ & -0.26  & -4.11 & -3.27& 15.2& 18.6\\\cline{1-6}
 \end{tabular}
 \vspace{0.8cm}
\caption{Parameters appearing in the form factors of the
$B_{S}\rightarrow D_{sJ}(2460)\ell\nu$}decay in a four-parameter
fit, for $M_{1}^2=15~GeV^2$, $M_{2}^2=6~GeV^2$ \label{tab:1}
\end{table}

 For completeness, let us discuss the heavy quark mass
limit of form factors. In this limit form factors for the
 $B_{s}\rightarrow D_{sJ}(2460)$ transition is calculated in \cite{21}. 
 In order to perform the heavy quark mass limit and estimate the
dependence of form factors $f_{V}$, $f_{0}$, $f_{+}$ and $f_{-}$ on
$y$ where
 \begin{equation}\label{18au}
 y=vv'=\frac{m_{B_{s}}^2+m_{D_{s}^{\ast}}^2-q^2}{2m_{B_{s}}m_{D_{s}^{\ast}}}
 \end{equation}
we follow the procedure as proposed in \cite{22} and evaluate the
sum rules at $q^2 =0 $ by taking $ m_{b}\rightarrow\infty$, with
$m_{c} = m_{b}/\sqrt{z}$ where z is fixed and is given by
$\sqrt{z}=y+\sqrt{y^2-1}$ at $q^2 = 0$. Here, v and v' are the four
velocities of $B_{s}$ and $D_{sJ}$ mesons, respectively. In the
$m_{b}\rightarrow\infty$ limit the Borel parameters $M_{1}^2$ and $
M_{2}^2$ take the form $M_{1}^2 =2T_{1}m_{b}$ and $ M_{2}^2=
2T_{2}m_{b}/\sqrt{z}$, where $T_{1}$ and $T_{2}$ are the new Borel
parameters. In this limit, the continuum thresholds $s_{0}$, and
$s_{0}'$ become
 \begin{equation}\label{19au}
 s_{0}=m_{b}^2+m_{b}\nu_{0},~~~~~~ s_{0}'=\frac{m_{b}^2}{z}+\nu_{0}'\frac{m_{b}}{\sqrt{z}}
 \end{equation}
 and the new integration variables $\nu$ and $ \nu'$ are defined as
\begin{equation}\label{20au}
s=m_{b}^2+m_{b}\nu,~~~~~~
s'=\frac{m_{b}^2}{z}+\nu'\frac{m_{b}}{\sqrt{z}}
\end{equation}
In the $m_{b}\rightarrow\infty$ limit leptonic decay constants
 $f_{B_{s}}$ and $f_{D_{sJ}}$ are rescaled as follows:
\begin{equation}\label{21au}
f_{B_{s}}=\frac{\hat{f}_{B_{s}}}{\sqrt{m_{b}}},~~~~~~~f_{D_{sJ}}~=\frac{\hat{f}_{D_{sJ}}}{\sqrt{m_{c}}}
\end{equation}
Taking into account the above-mentioned replacements,the sum rules
for the form factors \\
 $f_{V},~f_{0},~f_{+}$ and $f_{-}$ in the $
m_{b}\rightarrow\infty$ limit are given as
\begin{eqnarray}\label{22au}
f_{V}=-\frac{z^{1/4}(1+\frac{1}{\sqrt{z}})}{\hat{f}_{D_{sJ}}\hat{f}_{B_{s}}}e^{(\Lambda+\overline{\Lambda}/T)}\{
\frac{-3z^{2}}{(2\pi)^{2}(z-1)^{3}}\int_{0}^{\nu_{0}}d\nu\int_{0}^{\nu_{0}'}d\nu'(\nu-\nu')e^{\frac{-(\nu+\nu')}{2T}}
\nonumber\\
\theta(2y\nu\nu'-\nu^{2}-\nu'^2)+<\overline{\psi}\psi>[1+\frac{m_{0}^{2}}{T^{2}}(\frac{1}{24}+\frac{1}{12\sqrt{z}}
+\frac{\sqrt{z}}{12})]\},
\end{eqnarray}
\begin{eqnarray}\label{23au}
f_{0}=-\frac{z^{1/4}}{\hat{f}_{D_{sJ}}\hat{f}_{B_{s}}(1+\frac{1}{\sqrt{z}})}~e^{(\Lambda+\overline{\Lambda}/T)}~\{
\frac{-3z^{1/2}}{(4\pi)^{2}(z-1)}~\int_{0}^{\nu_{0}}d\nu\int_{0}^{\nu_{0}'}d\nu'(\nu-\nu')e^{\frac{-(\nu+\nu')}{2T}}
\nonumber\\
\theta(2y\nu\nu'-\nu^{2}-\nu'^2)+<\overline{\psi}\psi>[\frac{1}{2}+\frac{1}{2z}-\frac{1}{\sqrt{z}}
+\frac{m_{0}^{2}}{T^{2}}(\frac{-1}{24z^{3/2}}+\frac{1}{12\sqrt{z}}
-\frac{\sqrt{z}}{24})]\},
\end{eqnarray}
\begin{eqnarray}\label{24au}
f_{+}=-\frac{z^{1/4}(1+\frac{1}{\sqrt{z}})}{\hat{f}_{D_{sJ}}\hat{f}_{B_{s}}}e^{(\Lambda+\overline{\Lambda}/T)}\{
\frac{3z^{3/2}(\sqrt{z}-1)^4(\sqrt{z}+1)^2}{(4\pi)^{2}(z-1)^{5}}\int_{0}^{\nu_{0}}d\nu\int_{0}^{\nu_{0}'}d\nu'(\nu-\nu')e^{\frac{-(\nu+\nu')}{2T}}
\nonumber\\
\theta(2y\nu\nu'-\nu^{2}-\nu'^2)+<\overline{\psi}\psi>\frac{m_{0}^{2}}{T^{2}}[\frac{1}{12}+\frac{1}{24\sqrt{z}}
+\frac{\sqrt{z}}{24}]\},
\end{eqnarray}
\begin{eqnarray}\label{25au}
f_{-}=-\frac{z^{1/4}(1+\frac{1}{\sqrt{z}})}{\hat{f}_{D_{sJ}}\hat{f}_{B_{s}}}e^{(\Lambda+\overline{\Lambda}/T)}\{
\frac{3z^{3/2}(\sqrt{z}+1)^4(\sqrt{z}-1)^2}{(4\pi)^{2}(z-1)^{5}}\int_{0}^{\nu_{0}}d\nu\int_{0}^{\nu_{0}'}d\nu'(\nu-\nu')e^{\frac{-(\nu+\nu')}{2T}}
\nonumber\\
\theta(2y\nu\nu'-\nu^{2}-\nu'^2)-<\overline{\psi}\psi>\frac{m_{0}^{2}}{T^{2}}[\frac{1}{12}+\frac{1}{24\sqrt{z}}
+\frac{\sqrt{z}}{24}]\},
\end{eqnarray}
In deriving these results we take $T_{1}=T_{2}=T$ and the parameters
$\Lambda$ and $\overline{\Lambda}$ are obtained from two-point sum
rules that predicts $\Lambda=0.62~GeV$ \cite{23} and
$\overline{\Lambda}=0.86~GeV$ \cite{24}. Numerical analysis of the
above sum rules gives:
\begin{equation}\label{26au}
f_{V}(y_{max})=0.4,~~~~f_{~0}(y_{max})=0.031,~~~~~f_{+}(y_{max})=0.15,~~~~f_{-}(y_{max})=-0.27
\end{equation}
Where $y_{max}=1.30931$, which corresponds to $q^2=0$. When we
compare these results with the ones given in Eq.(\ref{17au}) at
 $q^2=0$, we
see that finite mass corrections are essential for $f_{V}(q^2)$ and
$f_{0}(q^2)$. Here we note that HQET limit for the transition form
factors for $B$ decays is discussed in \cite{25}.

\begin{figure}[p]
\begin{center}
\includegraphics[width=12cm]{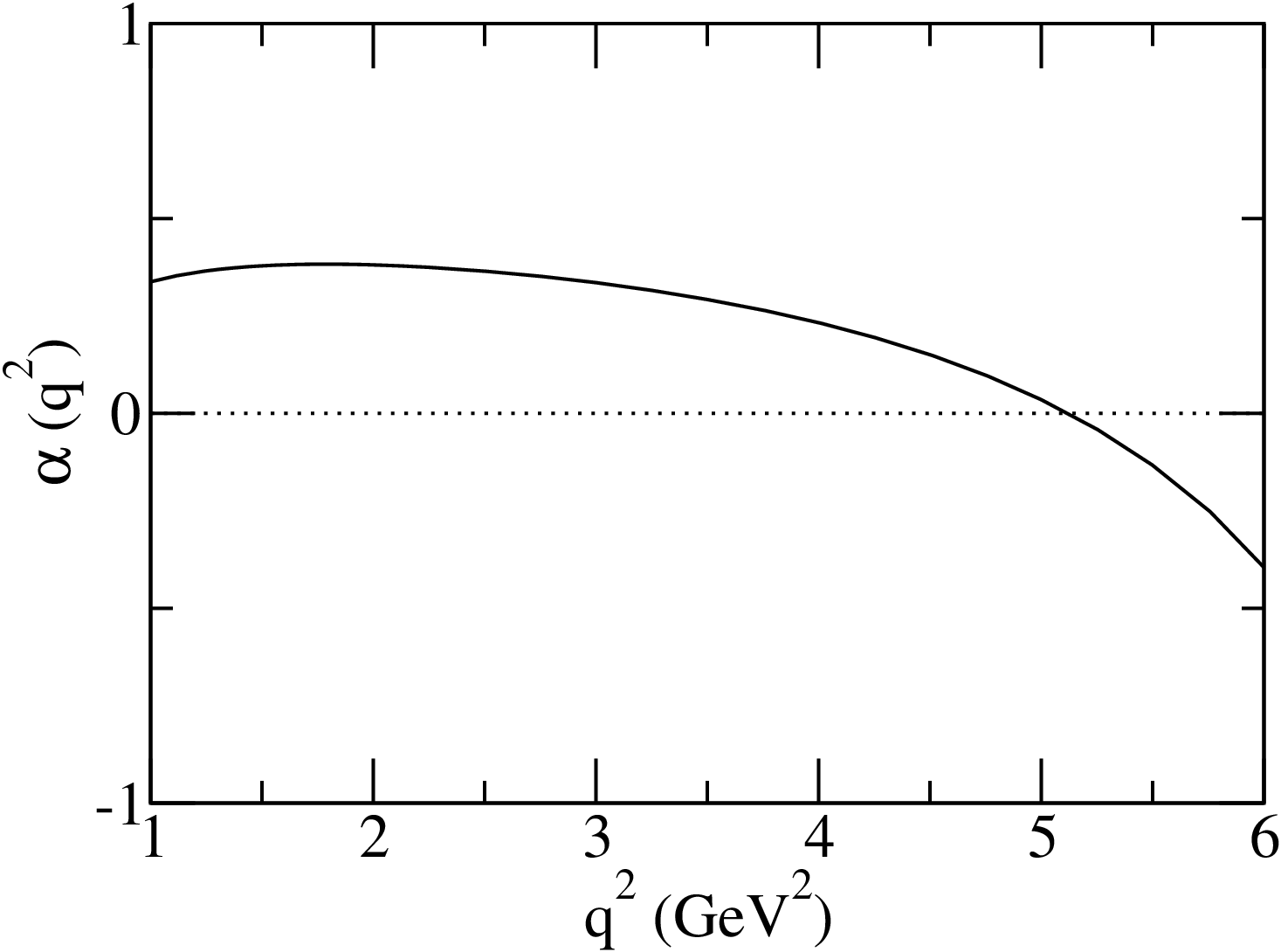}
\end{center}
\caption{The dependence of $\alpha$ on
 $q^2$.} \label{fig5}
\end{figure}

 For
$B_{s}\rightarrow D_{sJ}(2460)l\nu$ decay it is also possible to
determine the polarization of the $D_{sJ}(2460)$ meson. For this
aim we determine the asymmetry parameter $\alpha$, characterizing
the polarization of the $D_{sJ}(2460)$ meson,as
\begin{equation}\label{27au}
\alpha=2\frac{d\Gamma_{L}/dq^2}{d\Gamma_{T}/dq^2}-1
\end{equation}
 where $d\Gamma_{L}/dq^2$ and $d\Gamma_{T}/dq^2$ are differential  widths of the
decay to the states with longitudinal and transversal polarized
$D_{sJ}(2460)$ meson. After some calculations for differential
decay rates $d\Gamma_{L}/dq^2$ and $ d\Gamma_{T}/dq^2$ we get
\begin{eqnarray}\label{28au}
\frac{d\Gamma_{T}}{dq^2}=\frac{1}{8\pi^4m_{B_{s}^2}}\mid\overrightarrow{p'}\mid
G_{F}^2\mid V_{cb}\mid^2\{(2A+Bq^2)[\mid
f'_{V}\mid^2(4m_{B_{s}}^2\mid\overrightarrow{p'}\mid^2)+\mid
f'_{0}\mid^2]\}
\end{eqnarray}
\begin{eqnarray}\label{29au}
\frac{d\Gamma_{L}}{dq^2}&=&\frac{1}{16\pi^4m_{B_{s}^2}}|\overrightarrow{p'}|
G_{F}^2|V_{cb}|^2\left\{(2A+Bq^2)\left[\mid
f'_{V}\mid^2(4m_{B_{s}}^2\mid\overrightarrow{p'}\mid^2 \right. \right.
\nonumber \\ &&
+m_{B_{s}}^2\frac{\mid\overrightarrow{p'}\mid^2}{m_{D_{sJ}}}
(m_{B_{s}}^2-m_{D_{sJ}}^2-q^2))+\mid
f'_{0}\mid^2
\nonumber \\ &&
-\mid f'_{+}\mid^2\frac{m_{B_{S}}^2\mid\overrightarrow{p'}\mid^2}{m_{D_{sJ}}^2}(2m_{B_{S}}^2+2m_{D_{sJ}}^2
-q^2)-\mid
f'_{-}\mid^2\frac{m_{B_{S}}^2\mid\overrightarrow{p'}\mid^2}{m_{D_{sJ}}^2}q^2
\nonumber\\&&
-2 \left. \frac{m_{B_{S}}^2\mid\overrightarrow{p'}\mid^2}{m_{D_{sJ}}^2}(Re(f'_{0}
f_{+}^{'\ast}+f'_{0}
f_{-}^{'\ast}+(m_{B_{s}}^2-m_{D_{sJ}}^2)f'_{+}f_{-}^{'\ast}))\right]
\nonumber \\ && 
-2B\frac{m_{B_{S}}^2\mid\overrightarrow{p'}\mid^2}{m_{D_{sJ}}^2}
\left[\mid f'_{0}\mid^2+(m_{B_{s}}^2-m_{D_{sJ}}^2)^2\mid
f'_{+}\mid^2+q^4\mid f'_{-}\mid^2 \right.
\nonumber \\ && 
+2(\left. \left.m_{B_{s}}^2-m_{D_{sJ}}^2)Re(f'_{0}f_{+}^{'\ast})
+2q^2f'_{0}f_{-}^{'\ast}+2q^2(m_{B_{s}}^2-m_{D_{sJ}}^2)Re(f'_{+}f_{-}^{'\ast})
\right]\right\}
\nonumber \\ 
\end{eqnarray}
where\\
\begin{eqnarray}\label{30au}
\mid\overrightarrow{p'}\mid&=&\frac{\lambda^{1/2}(m_{B_{S}}^2,m_{D_{sJ}}^2,q^2)}{2m_{B_{S}}}\nonumber\\
A&=&\frac{1}{12q^2}(q^2-m_{l}^2)^2I_{0}\nonumber\\
B&=&\frac{1}{6q^4}(q^2-m_{l}^2)(q^2+2m_{l}^2)I_{0}\nonumber\\
I_{0}&=&\frac{\pi}{2}(1-\frac{m_{l}^2}{q^2})\nonumber\\
\end{eqnarray}
The dependence of the asymmetry parameter $\alpha$ on $q^2$ is shown
in Fig. \ref{fig5}. From this figure we see that asymmetry parameter
 $\alpha$ varies between -0.3 and 0.3 when $q^2$ lies in the region $m_{l}^2\leq q^2\leq 6~ GeV^2$.
 An interesting observation is that around $q^2=5.2 ~GeV^2$ the asymmetry parameter
changes sign. Therefore measurement of the polarization asymmetry
parameter
 $\alpha$ at fixed values of $q^2$ and determination of its
sign can give unambiguous information about quark structure of
$D_{sJ}$ meson.

 At the end of this section we would like to
present the value of the branching ratio of this decay.
 Taking into account the $q^2$ dependence of
the form factors and performing integration over $q^2$ in the limit
$m_{l}^2\leq q^2\leq(m_{B_{s}}-m_{D_{sJ}})^2$ and using the total
life-time $\tau_{B_{s}}=1.46\times10^{-12}s$ \cite{26} we get for
the branching ratio
\begin{eqnarray}\label{31au}
\textbf{\emph{B}}(B_s\rightarrow
D_{sJ}(2460)\ell\nu)\simeq4.9\times10^{-3}
\end{eqnarray}
which can be easily measurable at LHC.

In conclusion,the
semileptonic
   $B_{s}\rightarrow D_{sJ}(2460)\ell\nu$ decay is
  investigated in QCD sum rule method. The $q^2$ dependence of the
  transition form factors are evaluated. The dependence of the
  asymmetry parameter $\alpha$ on $q^2$ is investigated and
  the branching ratio is estimated to be measurably large at LHC.
\section{Acknowledgment}
  One of the authors (A.O) would like to thank TUBA-GEBIP for their financial support.
  
\clearpage
\newpage
\end{document}